\documentclass[aps,prl,twocolumn,superscriptaddress,amsmath,amssymb]{revtex4-2}
\usepackage{graphicx} 
\usepackage{epstopdf} 
\usepackage{dcolumn} 
\usepackage{xcolor} 
\usepackage{amssymb} 
\usepackage{hyperref} 
\usepackage[utf8]{inputenc} 
\usepackage[english]{babel} 
\usepackage[mathlines]{lineno} 
\usepackage{blindtext} 
\usepackage[export]{adjustbox}
\usepackage{tabularx}
\usepackage[separate-uncertainty=true, retain-zero-uncertainty=true, zero-decimal-to-integer=false, number-unit-product = \text{ }]{siunitx}
\usepackage[italic]{hepnames}
\usepackage{caption}
\usepackage{subcaption}
\usepackage{tikz}
\usepackage{booktabs}
\usepackage{orcidlink}

\graphicspath{{figures/}} 

\newcommand{\sigprob}{\ensuremath{\mathcal{P}_\text{FEI}}\xspace}
\newcommand{\mbc}{\ensuremath{M_{\textrm{bc}}}\xspace}
\newcommand{\deltae}{\ensuremath{\Delta E}\xspace}
\newcommand{\lumion}{\ensuremath{\SI{189}{\per\femto\barn}}\xspace}

\newcommand{\FEI}{\texttt{FEI}\xspace}

\newcommand{\mmiss}{\texorpdfstring{\ensuremath{M_{\mathrm{miss}}^{2}}}{MM²}}

\newcommand{\wlow}{\ensuremath{w_{\textrm{low}}}\xspace}
\newcommand{\whigh}{\ensuremath{w_{\textrm{high}}}\xspace}
\newcommand{\winc}{\ensuremath{w_{\textrm{incl.}}}\xspace}

\newcommand{\Ppislow}{\texorpdfstring{\HepParticle{\pi}{\mathrm{slow}}{}}{π_slow}}

\newcommand{\Dzbar}{\texorpdfstring{\HepAntiParticle{D}{}{\rm{0}}}{anti-D0}}
\newcommand{\Bz}{$B^0$}
\newcommand{\Bzbar}{\texorpdfstring{\HepAntiParticle{B}{}{\rm{0}}}{anti-B0}}

\renewcommand{\PUpsilonFourS}{\texorpdfstring{\ensuremath{\Upsilon\left(4S\right)}}{Y(4S)}}

\newcommand{\BzToDstEMuNu}{\texorpdfstring{\ensuremath{B^0 \to D^{*-}\ell\nu}}{B0->D*-(e+,mu+)nu)}}

\newcommand{\afb}{\ensuremath{A_{\mathrm{FB}}}}

\newcommand{\deltaA}{\ensuremath{\Delta
 \mathcal{A}_x}}

\newcommand{\sthree}{\ensuremath{S_3}}

\newcommand{\sfive}{\ensuremath{S_5}}

\newcommand{\sseven}{\ensuremath{S_7}}

\newcommand{\snine}{\ensuremath{S_9}}

\newcommand{\costhetaell}{\ensuremath{\cos\theta_{\ell}}}
\newcommand{\costhetav}{\ensuremath{\cos\theta_{V}}}
\DeclareSIUnit\clight{\text{\ensuremath{c}}}
\DeclareSIUnit\barn{b}
\DeclareSIUnit\b{\barn}
\DeclareSIUnit\fb{\femto\barn}
\DeclareSIUnit\invfb{\per\femto\barn}
\DeclareSIUnit\ab{\atto\barn}
\DeclareSIUnit\invab{\per\atto\barn}
\DeclareSIUnit\gev{\GeV}
\DeclareSIUnit\gevc{\GeV\per\clight}
\DeclareSIUnit\gevcc{\GeV\per\clight\squared}

\def\jpsill         {\ensuremath{\jpsi \to \ell^+\ell^-}\xspace}
\def\llgamma        {\ensuremath{\ep\en \to \ell^+\ell^-(\gamma)}\xspace}
\def\eell           {\ensuremath{\ep\en \to (\ep\en)\ell^+\ell^-}\xspace}

\input ./belle2sym.tex
\begin{document}



\pacs{
}

\title{Tests of light-lepton universality in angular asymmetries of $B^0 \to D^{*-} \ell \nu$ decays}
\noaffiliation
  \author{I.~Adachi\,\orcidlink{0000-0003-2287-0173}} 
  \author{K.~Adamczyk\,\orcidlink{0000-0001-6208-0876}} 
  \author{L.~Aggarwal\,\orcidlink{0000-0002-0909-7537}} 
  \author{H.~Aihara\,\orcidlink{0000-0002-1907-5964}} 
  \author{N.~Akopov\,\orcidlink{0000-0002-4425-2096}} 
  \author{A.~Aloisio\,\orcidlink{0000-0002-3883-6693}} 
  \author{N.~Anh~Ky\,\orcidlink{0000-0003-0471-197X}} 
  \author{D.~M.~Asner\,\orcidlink{0000-0002-1586-5790}} 
  \author{H.~Atmacan\,\orcidlink{0000-0003-2435-501X}} 
  \author{T.~Aushev\,\orcidlink{0000-0002-6347-7055}} 
% ...
\collaboration{The Belle II Collaboration}

\begin{abstract}
  
We present the first comprehensive tests of the universality of the light leptons in the angular distributions of semileptonic $B^0$-meson decays to charged spin-1 charmed mesons. We measure five angular-asymmetry observables as functions of the decay recoil that are sensitive to lepton-universality-violating contributions. We use events where one neutral \B is fully reconstructed in $\Upsilon\left(4S\right)\to\B\overline{B}$ decays in data corresponding to $189~\mathrm{fb}^{-1}$ integrated luminosity from electron-positron collisions collected with the Belle II detector. We find no significant deviation from the standard model expectations.

\end{abstract}

\maketitle


In the standard model, all leptons share the same electroweak coupling, a symmetry known as lepton universality (LU). Semileptonic \B-meson decays involving the quark transition $b \to c \, \ell \, \nu$ provide excellent sensitivity to potential new interactions that would violate this symmetry. Evidence for lepton-universality violation~(LUV) in the ratio of semileptonic decay rates to $\tau$ leptons relative to the light-leptons $\ell$, denoting electrons and muons, has been reported in the combination of results from the BaBar, Belle, and LHCb Collaborations~\cite{babar_1, babar_2, belle_hadronic, belle_semileptonic, belle_polarization, lhcb_1, lhcb_2, lhcb_3}. Recently, evidence of LUV between the light leptons at the four-standard-deviation level has been reported based on differences in their angular distributions in semileptonic \B decays to $D^*$ mesons~\cite{bobeth}. However, that analysis relied on a reinterpretation of Belle results \cite{Belle:2018ezy} that contained only one-dimensional projections of the multidimensional angular distributions that are needed to fully characterize such decays. We present the first light-lepton LU test using a complete set of angular-asymmetry observables chosen to suppress most theoretical and experimental uncertainties, thus optimizing sensitivity to LUV~\cite{bhattacharya}. This test is complementary to the branching-fractions-based LUV test in Ref.~\cite{rXemu}.

The semileptonic decay \BzToDstEMuNu{} is mediated in the standard model via $W$-boson exchange (charge conjugation is implied throughout). Due to the spin of the $D^{*-}$, which is reconstructed from its decay to a \Dzbar{} and a charged pion, the properties of the coupling and the spin of the virtual $W$ are encoded in angular distributions of the final-state particles. These can be fully characterized in terms of a recoil parameter and three helicity angles. The recoil parameter is defined as
\begin{linenomath}\begin{align}
    w & \equiv \frac{m_{B^0}^2 + m_{D^{*} }^2 - q^2}{ 2 m_B m_{D^{*} }},
\end{align}\end{linenomath}
where $m_{B^0}$ and $m_{D^{*}}$ are the known $B^0$ and $D^{*-}$ masses and $q$ is the four-vector of the momentum transferred to the dilepton system (natural units are used throughout). The helicity angles are defined as follows: $\theta_{\ell}$ is the angle between the direction of the charged lepton in the virtual $W$ frame and the $W$ in the $B^0$ frame, $\theta_{V}$ is the angle between the \Dzbar{} direction in the $D^{*-}$ frame and the $D^{*-}$ in the $B^0$ frame, and $\chi$ is the angle between the decay planes formed by the virtual $W$ and the $D^{*-}$ in the $B^0$ frame. Of these angles, only $\theta_{\ell}$ is correlated to lab-frame quantities.

The four-dimensional standard-model differential rate can be represented in terms of eight helicity amplitudes and as a function of $w$, \costhetaell, \costhetav, and $\chi$~\cite{Korner:1989qb, Gilman:1989uy}. It is possible to construct one- or two-dimensional integrals of these differential rates to isolate angular asymmetries that are sensitive to LUV, called \afb{}, \sthree{}, \sfive{}, \sseven{}, and \snine{} ~\cite{bhattacharya}. The forward-backward asymmetry \afb{} measures the tendency for the charged lepton to travel in the same direction as the virtual $W$. The \sthree{} and \snine{} asymmetries are sensitive to the alignment of the lepton and $D^*$ momenta, while \sfive{} and \sseven{} measure coupled alignments in the orientation of the $D$ with respect to the $D^*$. We redefine these asymmetries in terms of one-dimensional integrals
\begin{equation}
 \mathcal{A}_x(w) \equiv \left(\frac{\mathrm{d}\Gamma}{\mathrm{d}w}\right)^{-1} \left [ \int_{0}^{1} - \int_{-1}^{0}\right] \mathrm{d}x\frac{\mathrm{d}^2\Gamma}{\mathrm{d}w\mathrm{d}x},
\end{equation}
with $x=\cos\theta_{\ell}$ for \afb{}, $\cos2\chi$ for \sthree, $\cos\chi\cos\theta_V$ for \sfive, $\sin\chi\cos\theta_V$ for \sseven, and $\sin2\chi$ for \snine, as illustrated in the supplemental material \cite{supplemental}. The determination of each of the five asymmetries then reduces to measuring the signal yields $N^-_x$ with $x\in[-1,0)$ and $N^+_x$ with $x\in[0,1]$ after accounting for experimental effects such as resolution and detector acceptance. The asymmetries are then calculated as
\begin{equation}
    \mathcal{A}_x(w) = \frac{N^+_x(w) - N^-_x(w)}{N^+_x(w) + N^-_x(w)}.
\end{equation}
The differences between the angular asymmetries of electrons and muons,
\begin{equation}
    \Delta\mathcal{A}_x(w) \equiv \mathcal{A}^{\mu}_x(w) - \mathcal{A}^{e}_x(w),
\end{equation}
are sensitive to interactions that violate LU. Most experimental uncertainties cancel in the asymmetries $\mathcal{A}_x$, and standard-model contributions largely cancel in the asymmetry differences \deltaA, only arising from the differences in lepton masses. Therefore, the compatibility between measurements of the asymmetry differences \deltaA{} and their standard-model expectations is a powerful test of LU. To optimize sensitivity to extensions of the standard model~\cite{bhattacharya}, we measure these variables integrated over three $w$ ranges: the full phase-space (\winc), the low $w$ region (\wlow) from $1$ up to $1.275$, and the high $w$ region (\whigh) from $1.275$ to the kinematic endpoint at $1.503$. 

For each asymmetry $\mathcal{A}_x$ and $w$ range, we separate signal candidates into angular categories $+$ and $-$ based on the measured value of $x$. We determine the numbers of signal events $N^{\pm}_x$ with fits to distributions of \mmiss, the squared difference between the sum of the four-momenta of the colliding particles and the sum of the four-momenta of the reconstructed particles. The \mmiss{} distribution for correctly reconstructed signal events peaks near zero, while the distribution for backgrounds, which come mostly from $B\to D^{**}\ell\nu$ decays, does not peak. We correct these event numbers for detector efficiency, acceptance, and resolution effects determined from simulation in order to calculate unbiased asymmetries. 

Of the five asymmetries, only \afb{} and \sthree\ have been measured, but not differentially in $w$~\cite{bobeth,prim,chaoyi}. In the standard model or any extension thereof, \snine{} is always zero~\cite{bobeth}. Similarly, \sseven{} is always zero in the standard model and has reduced sensitivity to its extensions~\cite{bhattacharya}. In contrast, \afb{}, \sthree{}, and \sfive{} are highly sensitive to LUV via their asymmetry differences, which should show highly correlated deviations from the SM expectations in the case of new interactions. Therefore, correlated LUV signatures between the asymmetry differences can help to probe the nature of any new interactions. Therefore, the simultaneous determination of all asymmetries in different $w$ ranges provides a powerful test of LU and probes the nature of non-standard-model interactions.

We measure the asymmetries and their differences using a dataset corresponding to \lumion of electron-positron collisions at 10.58 GeV center-of-mass energy collected by the Belle II experiment between 2019 and 2021. We use the Belle II detector \cite{Abe:2010gxa} to reconstruct \PUpsilonFourS{} $\to\Bz\Bzbar$ decays. The detector consists of several nested subsystems in a cylindrical barrel, closed on either end with endcaps, arranged around the interaction region and nearly coaxial with the beams. The innermost subsystem is the vertex detector, composed of two layers of silicon pixels and four outer layers of silicon-strip detectors. During data collection for this analysis the outermost pixel layer only covered $15\%$ of the azimuth. Charged-particle trajectories (tracks) are reconstructed by a small-cell drift chamber (CDC) filled with a He $50\%$ and $\mathrm{C}_2\mathrm{H}_6$ $50\%$ gas mixture, which also provides a measurement of ionization energy-loss for particle identification. A Cherenkov-light imaging and time-of-propagation detector (TOP) provides charged pion and kaon identification information in the barrel region. This information is provided in the forward endcap by a proximity-focusing, ring-imaging Cherenkov detector with an aerogel radiator.
An electromagnetic calorimeter (ECL) consisting of CsI(Tl) crystals provides neutral-particle and electron identification information in the barrel and both endcaps.
All of the above subsystems are immersed in a uniform 1.5~T magnetic field that is nearly aligned with the electron beam and is generated by a superconducting solenoid situated outside the calorimeter.
The outermost subsystem, the \KL and muon identification detector, consists of scintillator strips in the endcaps and the inner part of the barrel, and resistive-plate chambers in the outer barrel, interleaved with iron plates that serve as a magnetic flux-return yoke.

We use Monte Carlo (MC) simulation to model the signal and backgrounds and to calculate reconstruction efficiencies. We use the software libraries \texttt{EvtGen}~\cite{evtgen}, \texttt{PYTHIA}~\cite{pythia8}, and \texttt{KKMC}~\cite{kkmc} to model particle production and decay, \texttt{PHOTOS}~\cite{PHOTOS} for photon radiation, and \texttt{GEANT4}~\cite{AGOSTINELLI2003250} for detector response. We overlay simulated beam-induced backgrounds on the simulated events~\cite{BeamBKG}. We simulate 900\invfb{} of \BzToDstEMuNu{} decays with the form factors of Refs.~\cite{Boyd:1994tt,Grinstein:2017nlq,Bigi:2017njr} and values determined by the measurements of Refs.~\cite{Belle:2018ezy}. We use the Belle~II analysis software, \texttt{basf2}~\cite{basf2, basf2-zenodo}, to reconstruct simulated and experimental data identically.

In each event, we use the full event interpretation (\FEI) algorithm~\cite{fei} to fully reconstruct one neutral \Bz, called the tag \Bz. The FEI reconstructs tag \Bz candidates in explicit hadronic decay cascades with no missing particles. Each tag \Bz candidate then consists of a collection of detected tracks and neutral energy depositions (clusters) and a hypothesis for the full \Bz decay cascade that produced them. We use three variables to select correctly reconstructed tags. The beam-constrained mass \mbc is calculated from the center-of-mass collision energy $\sqrt{s}$ and tag-$B$ momentum $\vec{p}_{B}$,
\begin{equation}
\mbc=\sqrt{\left(\sqrt{s}/2\right)^{2}- |{\vec{p}_{B}}|^{2}}.
\end{equation} The energy difference $\deltae = E_{B} - \sqrt{s}/2$ is the difference between the center-of-mass collision energy and tag-$B$ energy $E_B$. Finally, a tag-reconstruction confidence score, \sigprob, valued between zero and one, quantifies the agreement between the kinematic properties of the detected particles and the hypothesized decay cascade. 

Correctly and completely reconstructed \Bz candidates have $\mbc$ near the \Bz mass, $\deltae$ near zero, and $\sigprob$ near 1. We require that tag \Bz candidates satisfy $\mbc >5.27\gev$, $\deltae \in[-0.15, 0.1]\gev$, and $\sigprob>0.001$. If multiple tag \Bz candidates in an event pass these selections, we keep only the one with the highest value of \sigprob. 

In events with an identified tag \Bz candidate, we reconstruct $B^0 \to D^{*-}(\to \overline{D}{}^0\pi^-) \ell \nu$ candidates with \Dzbar{} decaying to \Kp{}\pim{}, \Kp{}\pim{}\pip{}\pim{}, \Kp{}\pim{}\piz{}, \Kp{}\pim{}\pip{}\pim{}\piz{}, \KS{}\pip{}\pim{}, \KS{}\pip{}\pim{}\piz{}, \KS{}\piz{}, or \Kp{}\Km{} final states. We require that all tracks originate from the vicinity of the interaction point. We require that each lepton candidate have a lab-frame momentum above $0.4\gev$, and a polar angle within the range $[0.22, 2.71]$~rad for electrons and $[0.4, 2.6]$~rad for muons, to ensure that suitable particle-identification information is available. Leptons are identified using the ratio of their likelihood to the sum of likelihoods for all charged-particle types. These likelihoods combine particle-identification information from the CDC, ECL, and, for muons, the TOP. We retain lepton candidates with a likelihood ratio above $0.9$, resulting in electron and muon identification efficiencies of 86\% and 89\%, respectively, and hadron misidentification rates of less than $1\%$ and $3\%$, respectively. We determine lepton-identification efficiencies and their uncertainties from auxiliary measurements in discrete intervals of lab-frame momentum, polar angle, and charge, using \jpsill, \llgamma, and \eell events. 

We reconstruct \piz{} candidates via decays to two photons. We identify photon candidates from ECL clusters unassociated with any matched tracks and with timing selections designed to minimize contamination from beam-induced backgrounds. We require that each \piz{} candidate have an invariant mass in the range $[0.120,0.145]\gev$, approximately four times the diphoton mass resolution. The \piz{} reconstruction and selection efficiency is approximately $0.3$.

We reconstruct \KS{} candidates via decays to two charged particles that are assigned the pion mass. We require that each \KS{} candidate has an invariant mass in the range $[0.3,0.7]\gev$ and that it can be fit to a common vertex that is displaced from the interaction point by at least one unit of the uncertainty of the vertex fit. We also require that the angle between the momentum of the \KS{} candidate and the displacement of the vertex from the interaction point be less than $0.64$~rad.

We require that the mass of each \Dzbar{} candidate is in the range $[1.85,1.88]\gev$, corresponding to approximately four times the peak resolution and centered on the known mass. We reconstruct $D^{*-}$ candidates by combining \Dzbar{} candidates with each of the remaining tracks, which we label \Ppislow, and require that the mass difference between the \Dzbar{} and $D^{*-}$ candidates is in the range $[0.143,0.148]\gev$, approximately four times its resolution. 

We combine $D^{*-}$ and lepton candidates to form signal \Bz candidates and use the \texttt{TreeFit}~\cite{treefit} algorithm to reject candidates that cannot be fit to consistent vertices. We then combine the signal and tag \Bz candidates and require that no additional tracks remain in the event and that the difference between the reconstructed energy and the collision energy is greater than $0.3\gev$, in order to reject hadronic decay backgrounds. If more than one candidate passes these requirements, we select only the one closest to expectation in the well-modeled quantity $|M(D^*)-M(D)|$.

\begin{figure}
        \centering
        \includegraphics[width=\columnwidth]{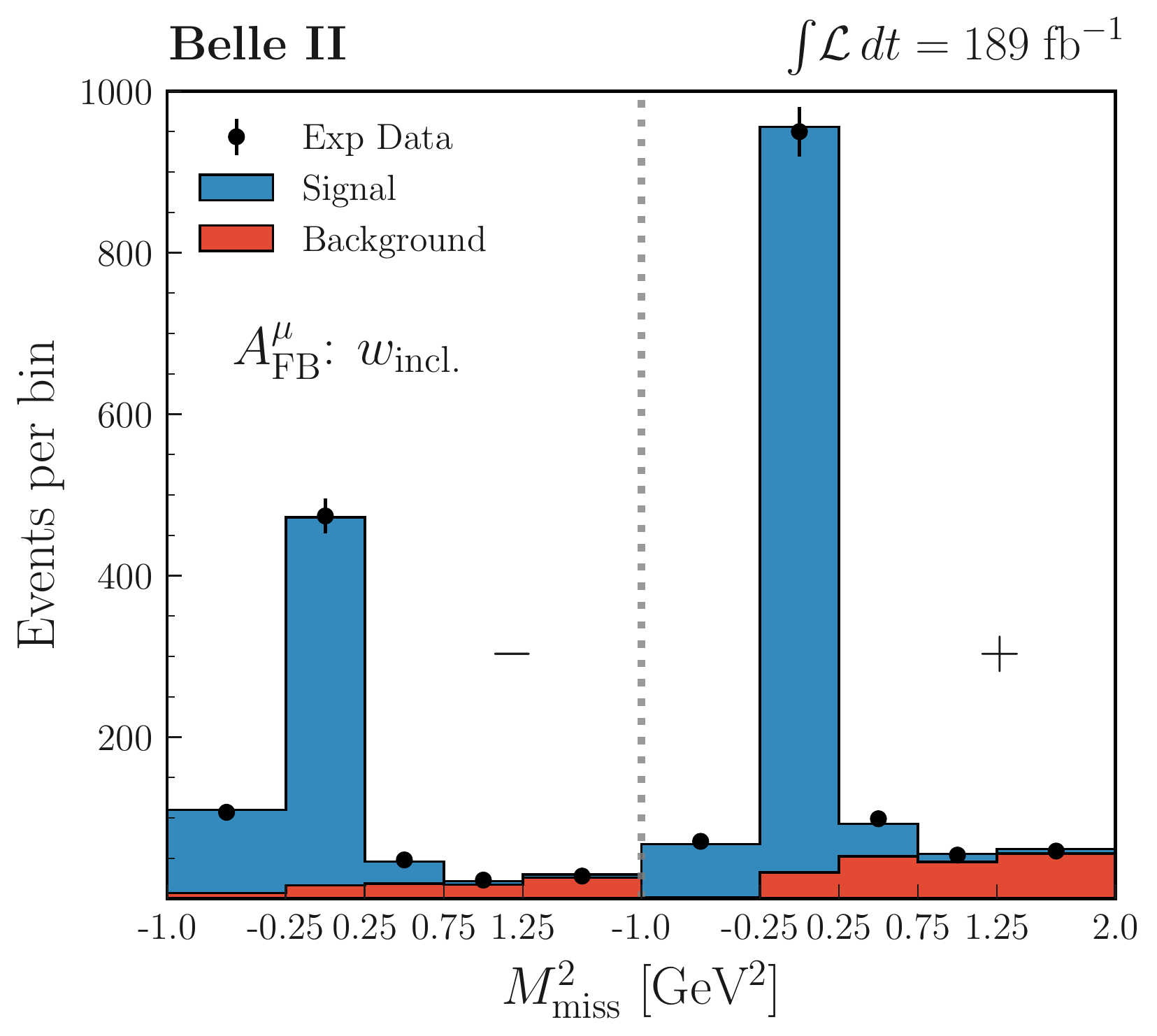}
	\caption{Muon-mode \mmiss{} distributions and fit results for $\cos\theta_\ell$ in the ranges $[-1,0)$ (left) and $[0,1]$ (right), corresponding to the $-$ and $+$ categories of $A_{\mathrm{FB}}^{\mu}$, in the full $w$ range (\winc).}
	\label{fig:fit}
\end{figure}

We obtain the signal yield for each combination of lepton flavor, $w$ interval, and angular category ($+$ or $-$ as defined for a particular asymmetry $\mathcal{A}_x$) using a binned maximum-likelihood fit to the distribution of \mmiss{}. The signal yield $N^{\pm}_x(w)$ and the background yield are unconstrained in the fit, allowing for an effective background-subtraction that removes dependence on the angular asymmetry of the backgrounds. We determine the shapes of the signal and background with simulation and choose a coarse binning to minimize dependence on resolution modeling. In Fig.~\ref{fig:fit} we show the two such independent fits that determine $N^+_x$ and $N^-_x$ for $x=\cos\theta_{\ell}$ in the muon mode and in the \winc bin. Together, these yields determine $\afb^{\mu}(\winc)$. We find 1617 (1639) signal events in the electron (muon) mode overall, with a variation of less than one event between variables. Of these, 803 (853) are in the \wlow range. 

We correct the fitted yields $N_x^{\pm}(w)$ for selection and detector acceptance losses using efficiency estimates from simulation. The efficiency ratios between the $+$ and $-$ categories are typically near $1.0$ but range up to nearly $1.4$ for $\afb$, largely due to the reduced momentum of leptons in the $-$ category, which are emitted opposite to the direction of the boost of the $B$ mesons, relative to the $+$ leptons, which are emitted in the direction of the boost. This lower momentum results in lower reconstruction and identification efficiencies. We further correct for migration of candidates between the $+$ and $-$ categories and different $w$ bins by inverting a detector-response matrix. This matrix is constructed from the conditional probabilities that events generated in a particular kinematic bin are reconstructed in each kinematic bin. For every variable and bin, the probability of reconstruction into the correct bin is above $0.86$.

The largest systematic uncertainty affecting the measurement is from the size of the simulated samples, which limits the precision of the bin-migration and efficiency corrections. We determine this uncertainty from the standard deviation of the results obtained by repeatedly resampling the simulated data with replacement and refitting. This uncertainty is approximately one-fourth to one-half of the statistical uncertainty, ranging in 0.010--0.025. We determine the uncertainties from other systematic effects by varying their contribution within their known uncertainties or bounds~\cite{Zyla:2020zbs} or from independent control data. Lepton-identification uncertainties mostly cancel in the asymmetries $\mathcal{A}$ and are at most $0.004$. The uncertainty on the reconstruction efficiency of $\pi_{\mathrm{slow}}$ also largely cancels and is negligible. Uncertainties from modeling of the background processes, such as $B \to D^{**} \ell \bar \nu_\ell$, are negligible due to fitting the backgrounds independently in $+$ and $-$ categories. The supplemental material contains a full list of all of the systematic uncertainties~\cite{supplemental}.

\begin{figure*}
    \centering
  \includegraphics[width=\textwidth]{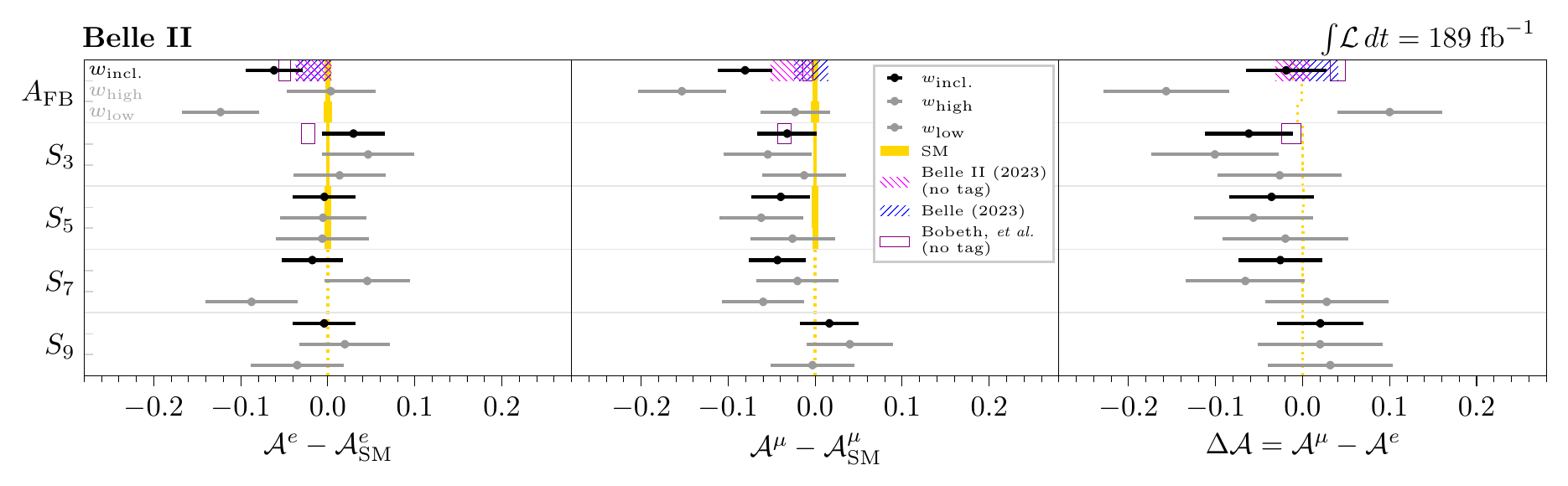}
  \caption{Observed asymmetries and their differences (points with error bars), one-standard-deviation bands from the Belle~\cite{prim} and Belle II~\cite{chaoyi} measurements (hatched boxes), calculations from Ref.~\cite{bobeth} based on a previous measurement from Belle~\cite{Belle:2018ezy}(empty boxes), and standard-model expectations (solid boxes). The standard-model expectation is drawn with a dashed line when its uncertainty is too small to display.
}
  \label{fig:results}
\end{figure*}

\begin{table}
\centering
\caption{Summary of the results and comparison with expectations. The measurement uncertainties are statistical and systematic, respectively.}
\label{tab:results_prl}
\begin{tabular}{llcc}
\toprule
                     Obs. &             $w$ bin &               $\Delta\mathcal{A}_x$ &   SM expectation\\
\midrule
 $\Delta A_{\mathrm{FB}}$ &    $w_\mathrm{low}$ &    $\phantom{-}0.099\pm0.056\pm0.020$ &  $-0.00104$\\
                          &   $w_\mathrm{high}$ &  $-0.168\pm0.068\pm0.024$ &   $-0.01133$\\
                          &  $w_\mathrm{incl.}$ &  $-0.024\pm0.043\pm0.016$ &   $-0.00566$\\
             $\Delta S_3$ &    $w_\mathrm{low}$ &  $-0.026\pm0.068\pm0.024$ &     $\phantom{-}0.00028$\\
                          &   $w_\mathrm{high}$ &  $-0.101\pm0.069\pm0.025$ &     $\phantom{-}0.00023$\\
                          &  $w_\mathrm{incl.}$ &  $-0.062\pm0.047\pm0.017$ &     $\phantom{-}0.00018$\\
             $\Delta S_5$ &    $w_\mathrm{low}$ &  $-0.019\pm0.068\pm0.024$ &     $\phantom{-}0.00027$\\
                          &   $w_\mathrm{high}$ &  $-0.055\pm0.065\pm0.023$ &     $\phantom{-}0.00107$\\
                          &  $w_\mathrm{incl.}$ &  $-0.035\pm0.046\pm0.016$ &     $\phantom{-}0.00049$\\
             $\Delta S_7$ &    $w_\mathrm{low}$ &   $\phantom{-}0.028\pm0.067\pm0.024$ &      $\phantom{-}0\phantom{.00000}$\\
                          &   $w_\mathrm{high}$ &  $-0.066\pm0.065\pm0.022$ &      $\phantom{-}0\phantom{.00000}$\\
                          &  $w_\mathrm{incl.}$ &  $-0.026\pm0.046\pm0.016$ &      $\phantom{-}0\phantom{.00000}$\\
             $\Delta S_9$ &    $w_\mathrm{low}$ &   $\phantom{-}0.032\pm0.067\pm0.024$ &      $\phantom{-}0\phantom{.00000}$\\
                          &   $w_\mathrm{high}$ &    $\phantom{-}0.020\pm0.068\pm0.024$ &     $\phantom{-}0\phantom{.00000}$\\
                          &  $w_\mathrm{incl.}$ &    $\phantom{-}0.020\pm0.047\pm0.017$ &     $\phantom{-}0\phantom{.00000}$\\
\bottomrule
\end{tabular}
\end{table}

Figure~\ref{fig:results} shows our measurements of the asymmetries and the LUV-sensitive differences and Table~\ref{tab:results_prl} shows the numerical values. The numerical values and full covariance matrices of the measured observables will be made available on HEPData \cite{hepdata}. These measurements are the first comprehensive tests of lepton universality in the angular distributions of semileptonic $B$ decays. We compare our measurements to predictions from Ref.~\cite{Bernlochner:2022ywh} and measurements from Refs.~\cite{bobeth,prim,chaoyi}. The results in Ref.~\cite{bobeth} are obtained in a slightly reduced $w$ range, $[1,1.5]$, which makes them not strictly comparable to the other results. However, the standard-model expectations in these two $w$ ranges differ only in the fourth decimal place. The results from Refs.~\cite{bobeth, chaoyi} derive from analyses without explicit reconstruction of the tag \B, resulting in lower statistical uncertainties relative to these results. 

To test agreement with the standard-model expectation \cite{Bernlochner:2022ywh}, we perform three different $\chi^2$ tests, accounting for the statistical and systematic covariances between all of the variables. Tests of the asymmetries $\mathcal{A}$ in the full $w$ range (\winc) yield $\chi^2/N_{\textrm{dof}}=14.6/10$ ($p=0.15$) and in $w$ subranges (\wlow, \whigh) yield $26.7/20$ ($p=0.14$). Tests of the LUV-sensitive asymmetry differences $\Delta\afb$, $\Delta\sthree$, and $\Delta\sfive$ in the \winc range yield $\chi^2/N_{\textrm{dof}}=2.0/3$ ($p=0.57$) and in $w$ subranges yield $10.2/6$ ($p=0.13$). Tests of the insensitive quantities $\Delta\sseven$ and $\Delta\snine$ in the \winc range yield $\chi^2/N_{\textrm{dof}}=0.6/2$ ($p=0.76$) and in $w$ subranges yield $1.5/4$ ($p=0.83$). Our results agree well with the standard-model expectations and provide no evidence for LUV.



This work, based on data collected using the Belle II detector, which was built and commissioned prior to March 2019, was supported by
Science Committee of the Republic of Armenia Grant No.~20TTCG-1C010;
Australian Research Council and research Grants
No.~DE220100462,
No.~DP180102629,
No.~DP170102389,
No.~DP170102204,
No.~DP150103061,
No.~FT130100303,
No.~FT130100018,
and
No.~FT120100745;
Austrian Federal Ministry of Education, Science and Research,
Austrian Science Fund
No.~P~31361-N36
and
No.~J4625-N,
and
Horizon 2020 ERC Starting Grant No.~947006 ``InterLeptons'';
Natural Sciences and Engineering Research Council of Canada, Compute Canada and CANARIE;
Chinese Academy of Sciences and research Grant No.~QYZDJ-SSW-SLH011,
National Natural Science Foundation of China and research Grants
No.~11521505,
No.~11575017,
No.~11675166,
No.~11761141009,
No.~11705209,
and
No.~11975076,
LiaoNing Revitalization Talents Program under Contract No.~XLYC1807135,
Shanghai Pujiang Program under Grant No.~18PJ1401000,
Shandong Provincial Natural Science Foundation Project~ZR2022JQ02,
and the CAS Center for Excellence in Particle Physics (CCEPP);
the Ministry of Education, Youth, and Sports of the Czech Republic under Contract No.~LTT17020 and
Charles University Grant No.~SVV 260448 and
the Czech Science Foundation Grant No.~22-18469S;
European Research Council, Seventh Framework PIEF-GA-2013-622527,
Horizon 2020 ERC-Advanced Grants No.~267104 and No.~884719,
Horizon 2020 ERC-Consolidator Grant No.~819127,
Horizon 2020 Marie Sklodowska-Curie Grant Agreement No.~700525 "NIOBE"
and
No.~101026516,
and
Horizon 2020 Marie Sklodowska-Curie RISE project JENNIFER2 Grant Agreement No.~822070 (European grants);
L'Institut National de Physique Nucl\'{e}aire et de Physique des Particules (IN2P3) du CNRS (France);
BMBF, DFG, HGF, MPG, and AvH Foundation (Germany);
Department of Atomic Energy under Project Identification No.~RTI 4002 and Department of Science and Technology (India);
Israel Science Foundation Grant No.~2476/17,
U.S.-Israel Binational Science Foundation Grant No.~2016113, and
Israel Ministry of Science Grant No.~3-16543;
Istituto Nazionale di Fisica Nucleare and the research grants BELLE2;
Japan Society for the Promotion of Science, Grant-in-Aid for Scientific Research Grants
No.~16H03968,
No.~16H03993,
No.~16H06492,
No.~16K05323,
No.~17H01133,
No.~17H05405,
No.~18K03621,
No.~18H03710,
No.~18H05226,
No.~19H00682, 
No.~22H00144,
No.~26220706,
and
No.~26400255,
the National Institute of Informatics, and Science Information NETwork 5 (SINET5), 
and
the Ministry of Education, Culture, Sports, Science, and Technology (MEXT) of Japan;  
National Research Foundation (NRF) of Korea Grants
No.~2016R1\-D1A1B\-02012900,
No.~2018R1\-A2B\-3003643,
No.~2018R1\-A6A1A\-06024970,
No.~2018R1\-D1A1B\-07047294,
No.~2019R1\-I1A3A\-01058933,
No.~2022R1\-A2C\-1003993,
and
No.~RS-2022-00197659,
Radiation Science Research Institute,
Foreign Large-size Research Facility Application Supporting project,
the Global Science Experimental Data Hub Center of the Korea Institute of Science and Technology Information
and
KREONET/GLORIAD;
Universiti Malaya RU grant, Akademi Sains Malaysia, and Ministry of Education Malaysia;
Frontiers of Science Program Contracts
No.~FOINS-296,
No.~CB-221329,
No.~CB-236394,
No.~CB-254409,
and
No.~CB-180023, and No.~SEP-CINVESTAV research Grant No.~237 (Mexico);
the Polish Ministry of Science and Higher Education and the National Science Center;
the Ministry of Science and Higher Education of the Russian Federation,
Agreement No.~14.W03.31.0026, and
the HSE University Basic Research Program, Moscow;
University of Tabuk research Grants
No.~S-0256-1438 and No.~S-0280-1439 (Saudi Arabia);
Slovenian Research Agency and research Grants
No.~J1-9124
and
No.~P1-0135;
Agencia Estatal de Investigacion, Spain
Grant No.~RYC2020-029875-I
and
Generalitat Valenciana, Spain
Grant No.~CIDEGENT/2018/020
Ministry of Science and Technology and research Grants
No.~MOST106-2112-M-002-005-MY3
and
No.~MOST107-2119-M-002-035-MY3,
and the Ministry of Education (Taiwan);
Thailand Center of Excellence in Physics;
TUBITAK ULAKBIM (Turkey);
National Research Foundation of Ukraine, project No.~2020.02/0257,
and
Ministry of Education and Science of Ukraine;
the U.S. National Science Foundation and research Grants
No.~PHY-1913789 
and
No.~PHY-2111604, 
and the U.S. Department of Energy and research Awards
No.~DE-AC06-76RLO1830, 
No.~DE-SC0007983, 
No.~DE-SC0009824, 
No.~DE-SC0009973, 
No.~DE-SC0010007, 
No.~DE-SC0010073, 
No.~DE-SC0010118, 
No.~DE-SC0010504, 
No.~DE-SC0011784, 
No.~DE-SC0012704, 
No.~DE-SC0019230, 
No.~DE-SC0021274, 
No.~DE-SC0022350; 
and
the Vietnam Academy of Science and Technology (VAST) under Grant No.~DL0000.05/21-23.

These acknowledgements are not to be interpreted as an endorsement of any statement made
by any of our institutes, funding agencies, governments, or their representatives.

We thank the SuperKEKB team for delivering high-luminosity collisions;
the KEK cryogenics group for the efficient operation of the detector solenoid magnet;
the KEK computer group and the NII for on-site computing support and SINET6 network support;
and the raw-data centers at BNL, DESY, GridKa, IN2P3, INFN, and the University of Victoria for offsite computing support.

\bibliography{references}


\end{document}